# Search For Gravity in Quantum Evolution

Aalok Pandya

Department of Physics, University of Rajasthan, Jaipur 302004, India.

The possibilities of curvature of space-time in the metric of quantum states are investigated. The curvature of the metric corresponding to a wave function of Hydrogen atom is determined. Also, Einstein tensor is described for a given quantum state.

______________________

**E-mail*: belagem@datainfosys.net

# 1. Introduction

The possible applications of the metric of quantum states in the configuration space calculated by Aalok et al [1], are investigated in this letter. Physicists often attempt to describe the gravity of stars, galaxies and the Universe. Now, it is worth asking what is the curvature of the space-time corresponding to an object like Hydrogen atom. How the quantum understanding of Hydrogen atom accommodates with the general theory of relativity? The present study is an attempt to investigate this aspect of gravitation in the light of study of quantum evolution [1].

In the construction of classical theory of gravitation in General Relativity, we recognize certain significant stages viz., (i) definition of manifold (ii) formulation of metric tensor (iii) description of parallel transport and geodesics, and lastly (iv) formulation of curvature tensor and Einstein's field equation [2]. In the studies of quantum evolution also, all these ideas pertaining to gravity have been revisited. Although, they are not identified exactly with quantum gravity but they are coming up nevertheless in the same manner in which general theory of relativity was developed. However, it cannot be called quantization of gravity, but it does lead to better understanding of quantum nature of gravity. During past few years many geometric concepts relevant to understanding of quantum nature of gravity have been discovered in the study of quantum evolution, for example concepts like 'length', 'distance' and 'geometric phases', have been described in these studies [3-4]. The metric of quantum state space has been formulated [3-5]. Also, the metric of quantum states has been described in the configuration space [1]. Manifold and spherical manifold has been described in the study of quantum evolution [5-6] and also, parallel transport and geodesic formulation has been explored [7]. Presently, in search of quantum nature of gravity various perturbative and nonperturbative approaches are in progress such as: nonperturbative approach to quantum gravity in search of quantum structure of space-time, initiated by Ashtekar and others [8], theory of super gravity [9], and Hawking's path of bridging classical and quantum theories of gravitation [10-11]. And many diverse and non-standard attempts are going on. And yet there is scope for exploring the nature of geometric and gravitational properties associated with physical objects identified by their respective quantum states. In this letter we analyze and explore the possible applications of the metric tensor [1] in the premise of General Theory of Relativity. We believe that these studies of quantum evolution will lead us to a better understanding of quantum nature of gravity.

## 2. Curvature in quantum evolution

In the study of quantum evolution the metric of quantum states in the configuration space has been obtained [1] as:

$$g_{\mu\nu} = \text{Re}\left[\left(\frac{\partial \Psi}{\partial x_\mu}\right)\left(\frac{\partial \Psi}{\partial x_\nu}\right)\right]. \tag{1}$$

And the corresponding invariant $ds$ is given by [1]:

$$ds^2 = \text{Re}\left[\left(\frac{\partial \Psi}{\partial x_\mu}\right)\left(\frac{\partial \Psi}{\partial x_\nu}\right)\right]dx^\mu dx^\nu. \tag{2}$$

If we choose to write this line element in a specific fashion, e.g. in the polar co-ordinates $(r,\theta,\varphi,t)$ with signature (+, +, +, -), the aforesaid expression appears as:

$$ds^2 = \text{Re}\left[\left(\frac{\partial \Psi}{\partial r}\right)\left(\frac{\partial \Psi}{\partial r}\right)\right]dr^2 + \text{Re}\left[\left(\frac{\partial \Psi}{\partial \theta}\right)\left(\frac{\partial \Psi}{\partial \theta}\right)\right]d\theta^2 + \text{Re}\left[\left(\frac{\partial \Psi}{\partial \varphi}\right)\left(\frac{\partial \Psi}{\partial \varphi}\right)\right]d\varphi^2 - \text{Re}\left[\left(\frac{\partial \Psi}{c\partial t}\right)\left(\frac{\partial \Psi}{c\partial t}\right)\right]c^2 dt^2. \tag{3}$$

We can compare this metric with any other generalized metric defined in the same coordinates and with the same signature. Thus, we examine and analyze its relevance by comparing it with Robertson-Walker line element, which represents a metric evolving with time:

$$ds^2 = S^2(t)\left[\frac{dr^2}{1-kr^2} + r^2(d\theta^2 + \sin^2\theta d\varphi^2)\right] - c^2 dt^2. \tag{4}$$

The comparison of $g_{\mu\nu}$ from (3) with (4) implies:

$$-c^2 = -c^2 \, \text{Re}\left[\frac{1}{c^2}\left(\frac{\partial \Psi}{\partial t}\right)\overline{\left(\frac{\partial \Psi}{\partial t}\right)}\right], \tag{5}$$

and 
$$\frac{S^2(t)}{1-kr^2} = \text{Re}\left[\left(\frac{\partial \Psi}{\partial r}\right)\overline{\left(\frac{\partial \Psi}{\partial r}\right)}\right]; \tag{6}$$

which implies 
$$k = \frac{1}{r^2}\left[1 - \frac{S^2(t)}{\text{Re}\left[\left(\frac{\partial \Psi}{\partial r}\right)\overline{\left(\frac{\partial \Psi}{\partial r}\right)}\right]}\right]. \tag{7}$$

Thus we can calculate curvature $k$, which obviously depends upon the nature of the wave function $\Psi$ and the scale factor $S(t)$.

Also we find that 
$$r^2 S^2(t) = \text{Re}\left[\left(\frac{\partial \Psi}{\partial \theta}\right)\overline{\left(\frac{\partial \Psi}{\partial \theta}\right)}\right], \tag{8}$$

and 
$$r^2 \sin^2\theta \, S^2(t) = \text{Re}\left[\left(\frac{\partial \Psi}{\partial \varphi}\right)\overline{\left(\frac{\partial \Psi}{\partial \varphi}\right)}\right]. \tag{9}$$

To illustrate an example, we calculate curvature of the metric corresponding to a wave function $\Psi$ of Hydrogen atom, as described in [1] by:

$$ds^2 = \left[C_1^2 \sin^2\theta \cos^2\varphi \left(e^{-\frac{r}{a_0}}\right)\left(1-\frac{r}{2a_0}\right)^2 \cos 2\omega t\right]dr^2 + \left[C_1^2 r^2 \cos^2\theta \cos^2\varphi \left(e^{-\frac{r}{a_0}}\right)\cos 2\omega t\right]d\theta^2$$

$$+ \left[C_1^2 r^2 \sin^2\theta \sin^2\varphi \left(e^{-\frac{r}{a_0}}\right)\cos 2\omega t\right]d\varphi^2 - \left[C_1^2 \omega^2 r^2 \sin^2\theta \cos^2\varphi \left(e^{-\frac{r}{a_0}}\right)\cos 2\omega t\right]dt^2$$

$$\tag{10}$$

The comparison of the equation (4) with equation (10) gives

$$\frac{S^2(t)}{(1-kr^2)} = C_1^2 \sin^2\theta \cos^2\varphi \left(e^{-r/a_0}\right)\left(1-\frac{r}{2a_0}\right)^2 \cos 2\omega t \,;$$

so that $k = \dfrac{1}{r^2}\left[1 - \dfrac{S^2(t)}{C_1^2 \sin^2\theta \cos^2\varphi \left(e^{-\frac{r}{a_0}}\right)\left(1-\dfrac{r}{2a_0}\right)^2 \cos 2\omega t}\right]$, (11)

with $S^2(t) = C_1^2 \cos^2\theta \cos^2\varphi \left(e^{-\frac{r}{a_0}}\right)\cos 2\omega t$. (12)

Therefore $k = \dfrac{1}{r^2}\left[1 - \dfrac{\cot^2\theta}{\left(1-\dfrac{r}{2a_0}\right)^2}\right]$. (13)

We notice that the expression of $k$ in equation (13) is singular at $r = 0$; $\theta = 0$; and $r = 2a_0$. But the singularities at $r = 0$ and $\theta = 0$ are the coordinate singularities and the only real physical singularity exists at $r = 2a_0$ i.e. at the distance twice of Bohr's radius.

### 3. Einstein tensor and Einstein's equation

Now we reformulate Einstein tensor and Einstein's field equation with the help of the metric of quantum states. Here we cautiously specify that this approach could be only one particular type of treatment with Einstein's field equation in this new scenario.

We can calculate curvature tensor $R_{\mu\nu}$ and hence Einstein's tensor $R_{\mu\nu} - \frac{1}{2} R g_{\mu\nu}$. Now we calculate the energy-momentum tensor $T_{\mu\nu}$ with the specific form:

$$T_{\mu\nu} = \rho\, v_\mu v_\nu = \rho c^2 \frac{dx_\mu}{ds}\frac{dx_\nu}{ds} \tag{14}$$

where $ds \approx cdt$.

To change this into quantum mechanical expression we make following transformation:

With
$$P_\mu \to -i\hbar \frac{\partial}{\partial x_\mu}, \tag{15}$$

we get
$$v_\mu^2 = \frac{\hbar^2}{m^2}\left\langle \frac{\partial \Psi}{\partial x_\mu}\bigg|\frac{\partial \Psi}{\partial x_\mu}\right\rangle, \tag{16}$$

and similarly
$$v_\nu^2 = \frac{\hbar^2}{m^2}\left\langle \frac{\partial \Psi}{\partial x_\nu}\bigg|\frac{\partial \Psi}{\partial x_\nu}\right\rangle. \tag{17}$$

Therefore
$$v_\mu v_\nu = \frac{\hbar^2}{m^2}\left[\left\langle \frac{\partial \Psi}{\partial x_\mu}\bigg|\frac{\partial \Psi}{\partial x_\mu}\right\rangle \left\langle \frac{\partial \Psi}{\partial x_\nu}\bigg|\frac{\partial \Psi}{\partial x_\nu}\right\rangle\right]^{\frac{1}{2}}. \tag{18}$$

Alternatively, with the knowledge of metric tensor in quantum mechanical form we can express energy momentum tensor as a metric derivative of action S, which can be calculated for a given state as:

$$T_{\mu\nu} = \frac{\delta}{\delta h_{\mu\nu}} S \tag{19}$$

Thus the quantum mechanical expression of the reformulated Einstein's field equation is;

$$R_{\mu\nu} - \frac{1}{2} R g_{\mu\nu} = -\frac{8\pi G}{c^4} \left( \frac{\rho \hbar^2}{m^2} \right) \left[ \left\langle \frac{\partial \Psi}{\partial x_\mu} \middle| \frac{\partial \Psi}{\partial x_\mu} \right\rangle \left\langle \frac{\partial \Psi}{\partial x_\nu} \middle| \frac{\partial \Psi}{\partial x_\nu} \right\rangle \right]^{\frac{1}{2}} \quad (20)$$

However, we are aware that this is not an eigen-equation and the terms occurring in this equation such as $\rho$ and $m$ are not quantum mechanical observable. But there are still two advantages of this formalism viz.

(i) *We can ascertain the Einstein tensor in terms of universal gravitational constant, by calculating the right hand side of equation* (20) *which yields a numeric figure for a specific wave function under consideration.*

(ii) *Various cosmological models described for specific quantum states can be tested with this reformulated equation.*

## 4. The Metric and the wave function of the Universe

Lastly, what we point out here is related with the idea of the wave function of the Universe [11]. The wave function of the Universe as formulated by Hartley and Hawking is metric dependent function:

$$\Psi_\pm[h_{ij}, \Phi] = \int_{c_\pm} d[g_{\mu\nu}] d[\Phi] \exp(-\tilde{I}[g_{\mu\nu}]) \quad (21)$$

The metric considered in this wave function is classical one. We want to point out here that for the new born Universe (early Universe) just after big-bang, the very initial wave function at time t≠0, could be given as a function of coordinates with respect to origin. And thereafter with evolution in time, metric $g_{\mu\nu}$ can be calculated with the help of evolving wave function. Thus, the very initial wave function of the Universe has to be coordinate dependent and thereafter wave function of the Universe can be expressed as a function of metric $g_{\mu\nu}$.